\documentclass[11pt,letterpaper]{article}
\usepackage[utf8]{inputenc}
\usepackage[left=1in,right=1in,top=1in,bottom=1in]{geometry}
\usepackage{amsmath}
\usepackage{amsfonts,amssymb} 
\usepackage{lmodern}
\usepackage{siunitx}
\usepackage{physics}
\usepackage{comment}
\usepackage{cite}
\usepackage[affil-it]{authblk}
\usepackage{rotating}
\usepackage{pdflscape}
\usepackage{fancyhdr}
\usepackage{multirow}
\usepackage{caption}
\usepackage{subcaption}
\usepackage{epsfig,epstopdf}
\usepackage{graphicx,color}
\usepackage[colorinlistoftodos,prependcaption,textsize=tiny]{todonotes}
\usepackage{lineno}
\usepackage{enumitem}
\usepackage{xcolor}

\newenvironment{Abstract}{\begin{quotation} \begin{center}
                       ABSTRACT
     \end{center}\bigskip  }{\end{quotation}}

\newcommand\snowmass{\begin{center}\rule[-0.2in]{\hsize}{0.01in}\\\rule{\hsize}{0.01in}\\
\vskip 0.1in Submitted to the  Proceedings of the US Community Study\\ 
on the Future of Particle Physics (Snowmass 2021)\\ 
\rule{\hsize}{0.01in}\\\rule[+0.2in]{\hsize}{0.01in} \end{center}}

\title{\bf Integration and Packaging}

\author[1]{S.M. Mazza\footnote{Author contact email: simazza@ucsc.edu}}
\author[2]{R. Lipton}
\author[3]{R. Patti}
\author[4]{R. Islam}
\affil[1]{SCIPP, University of California Santa Cruz, Santa Cruz (CA) 95064, US}
\affil[2]{Fermilab, Batavia(IL) 60510, US}
\affil[3]{NHanced semiconductors, Inc}
\affil[4]{Cactus materials, Inc}

\begin{document}

\maketitle

 \begin{Abstract}
\noindent Vertically integrated (3D) combinations of sensors and electronics provide the ability to fabricate small, fine pitch pixels with very small total capacitance monolithically integrated with complex circuitry. 
The small capacitance enabled by the fine pixel pitch, low interconnect capacitance, and very short signal path available in 3D hybrid bonding, provide an excellent signal to noise ratio with moderate power consumption. 
This combination enables fabrication of integrated sensors and electronics with both excellent position and time resolution. 
In this white paper, a discussion will be presented on the advantages of 3D integration, ongoing projects, and prospects in high energy physics and beyond.
\end{Abstract}

\snowmass

\setcounter{page}{1}

\section*{Executive summary}
In the past years,  high-energy physics experiments have been mostly relying on bump bonding for high density pixel to ASIC connection. 
The bump bonding technology was proven to be reliable and is currently used in many large silicon detector systems; however, it is known to have several limitations. It can be applied only down to 20-50~$\mu m$ of pitch and has yield issues for finer connections. 
Furthermore, the solder balls used for the connection inevitably increase the input capacitance to the amplifier and hence the noise. 
In order to provide a planar connection to a bump-bonded sensor, either an interposer is required or the sensor/chip needs to have suitable side extensions for external connections with wirebonds.
In terms of mechanical properties, the resulting connection is subject to thermal stress since it involves different materials, and repeated expansion and contraction of the assembly at higher or lower temperatures, respectively, can result in loss of electrical connection. The components' minimum thickness is also limited in practice, since in very thin silicon substrates, bending and mechanical stress can impact the feasibility of the procedure and decrease the yield of the bump bonding. This is particularly challenging in the case of larger silicon sensors. 

\begin{figure}[htbp]
\centering
\includegraphics[width=0.7\textwidth]{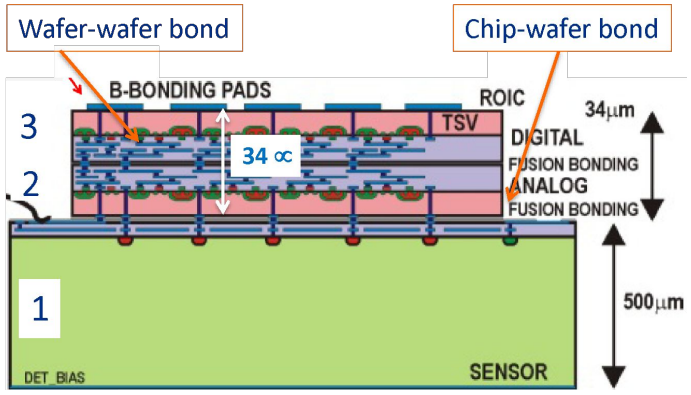}
\caption{Example of 3D integration of sensor and readout chip.}
\label{fig:IPDRvolt}
\end{figure}

The introduction of advanced sensor packaging may solve many of these issues, allowing for the improvement of performance, yield and processing.
3D integration is a widespread approach in industry, enabling tight packaging of sensor and readout chip. Furthermore it allows to stacks multiple chips in a single monolithic device. 
An example of 3D integration between sensor and chip is shown in Fig.~\ref{fig:IPDRvolt}.
There are several technologies available for 3D integration: the Direct Bonding Interconnect (DBI) is the most widely accepted and is implemented as silicon covalent oxide bonding or copper diffusion bonding.
The process can be done for a wafer to wafer (w2w) or die to wafer (d2w) assembly.
Through Silicon Via (TSV) connections allows to have multiple planes stacked and connected with external connections at top or bottom contacts, without the need of extensions or silicon interposers.
There are several advantages in using 3D integration in sensor to chip connection, a more exhaustive explanation is provided in Section~\ref{sec:advantages}:
\begin{itemize}
\item Less space: 3D chips can be multi-layered and do not need extension or interposers for external connections. Reduction of single layer thickness, after integration all supports can be removed
\item Very fine pitch bonding: down to a few micrometers
\item Better connections: faster and shorter than in circuit boards, with reduced dissipated power
\item Better performance: reduced input capacitance and lower noise
\item Layered design and integration: e.g. sensor + analog electronics + digital electronics, each layer can be manufactured by different producers
\item Reduced thermal stress and increased heat dissipation since material is homogeneous, also increased robustness
\end{itemize}

Advanced packaging and wafer to wafer bonding would facilitate several applications both in HEP and outside of HEP. A full description is provided in Section~\ref{sec:applications}, here a summary list follows: 4D tracking with high-granularity low-gain avalanche detectors, 3D integrated silicon photomultipliers with advanced signal processing, chip tile arrays assembled at wafer level with no edges, small pixels connection (10 ~$\mu m$ or less) that reduces the input capacitance of thin sensors (e.g. thin LGADs) increasing both space and time precision, double sided connection for LGADs to have readout on both sides, stacked 3D integrated chip, 3D network of algorithm cells for advanced pattern recognition, zero-mass trackers with very thin and highly populated layers, sensor stacks for high energy X-ray detection and decay detection. Finally, it will advance the knowledge in substrate engineering - useful, among other examples, in the fabrication of buried p-n layers for LGADs, necessary for the production of DJ-LGADs and buried-junction LGADs.

Therefore advanced packaging of electronics and sensors provides a variety of technologies that can meet the needs of future particle physics experiments. 
Combining these capabilities with silicon technologies developed for HEP, such as LGADs and active edge sensors, will allow the design of sophisticated detector systems that can meet the increasing challenges of next generation experiments. 
The collaboration (already ongoing, see section~\ref{sec:results}) between research groups and industry with established expertise in advanced packaging is crucial for the successful introduction of this technology in the research community (see section~\ref{sec:future}).

\section{Introduction}
Designs of modern silicon-based detector systems used for tracking and calorimetry are, 
to a large extent, dominated by limitations imposed by system packaging. Challenges for sensors and their associated readout electronics include 
device cooling and thermal interfaces, power dissipation, inter-device and off-detector 
communication, and low mass mechanical supports. There have been major advances in these 
areas that have enabled e.g. the large-scale silicon trackers in CMS and ATLAS and the silicon-based calorimetry in CMS. These include liquid CO$_2$-based cooling, extensive use of carbon fiber, increasingly complex hybrid and printed circuit board technology, and the availability of commercial CMOS processes (120-65 nm) in ASIC production for HEP experiments. Thus, state-of-the art packaging and integration technologies now available can significantly extend the reach and effectiveness of future detectors, enabling lower mass, finer pitch, and lower noise systems.

Recent HEP experiments have been mostly relying on bump bonding, also known as flip-chip bonding, for high-density pixel to ASIC interconnections. This connection technology was proven to be reliable; however it is know to have several limitations, both electrical and mechanical.
It provides good results down to the 50~$\mu$m scale, but with issues of yield arising at lower scale.
However, as the sensor technology reaches increasingly high granularity, it is necessary to develop a high-density interconnect process with the readout electronics that can be scaled down to even finer pitches.
While 3D integration of sensor and readout electronics is common in communication, computing and medical industries, this option is currently not available in the international HEP community. Such technologies for high-density interconnections would be of extreme interest for future HEP experiments. 
They will allow for finer pitch detectors with increased performance and mechanical stability, providing a high density interconnect between sensor and electronics with a readout pitch that can be lower than 10~$\mu$m.
This is crucial in many applications, for example for particle detection very close to the interaction point.

The following paper presents various advantages of 3D integration, ongoing projects, and prospects for 3D integration of silicon sensors in high energy physics and beyond.

\section{Advantages of advanced packaging}
\label{sec:advantages}

\subsection{Footprint}
Devices that share a package take up less space than if they were packaged separately. A multi-layer 3D chip may be no larger than a single traditional 2D chip, but it contains additional capabilities in each layer. Thus this design offers appreciable size reduction.
Furthermore the single components do not need a support wafer after integration, allowing for very thin layers of detectors and readout.

\subsection{Speed}
Devices assembled in 3D are much closer together than chips on a circuit board. The vertical 3D connections in the device stack are only a fraction of the length of circuit board wires. Shorter distances allow electronic signals to travel more quickly from one component to another, so a 3D assembly can operate faster than an equivalent circuit board. Past work has shown an improvement in latency or speed by a factor of 3-5. 

\subsection{Performance}
3D integration drastically lowers the parasitics of the interconnect wiring. TSVs typically have only 2-3~fF of capacitance and less than 3~$\Omega$ of resistance. In sensor applications, using DBI rather than micro-bumps has been shown to reduce noise and improve signal gain by 30\% or more. 

\subsection{Power}
The shorter connections in 3D assemblies also save significant power. Today, most of the power expended in systems is consumed by driving signals in wire, whether on-chip or across circuit boards. Additionally, when signals are external to a packaged device, the devices must have protection on the signal drivers and receivers to avoid ESD (electrostatic discharge) damage. ESD protection adds capacitance, and thus consumes more energy, when signals are driven in the interconnect wiring. 3D integration eliminates the need for ESD protection on the wiring between chips. Past work has shown 90\% or more power reduction in the energy consumed by the interconnect.

\subsection{Heterogeneous Integration}
The devices in a 3D assembly are manufactured separately, so they can vary widely. Each device may come from a different supplier, be built in a different process, incorporate different materials, etc. The components may differ in scale, voltage, dimensions, and any number of requirements. The most suitable technology option for a given application can be chosen for each functional layer separately.

\subsection{Robustness}
3D assembled layers will interface with one another through DBI connections, which have very few mechanical weaknesses. In fact, 3D devices display robustness similar to a counterpart monolithic 2D part before packaging. However, 2D devices need to interface through wire bonding or solder bumped connections, package-to-circuit board soldering, and the circuit board itself, all of which have mechanical weaknesses. Because 3D technology greatly reduces the number of packages involved in a system, it is better suited to withstand harsh environments than traditional circuit board collections of packaged 2D counterparts.


\section{Commercial Availability}
\label{sec:companies}
Variants of the 3D integration technology originally developed by Ziptronix as Direct Bond Interconnect (DBI) have now been adopted by an number of foundries under the generic term "Hybrid Bonding". 
This technology is now almost universally used for cell phone image sensors, most notably by Sony which licensed the process in 2015. Hybrid bonding is the process of bonding pairs of devices (wafer to wafer or chip-to-wafer) to achieve very fine pitch and robust interconnection. 
In addition to the hybrid bonding, connections must be extracted from the device layers which are buried after the bonding process.
This is usually done using Through-Silicon-Vias (TSVs) which can be inserted as part of the foundry process. 
Currently, TSVs are available to HEP from a few foundries~\footnote{Global Foundries, Skywater, Tower/Jazz, and MIT-LL}, with several others providing TSV insertion after the CMOS wafer is fully processed. Hybrid bonding itself has wider availability\footnote{NHanced, Cactus, Teledyne/Dalsa, IZM Dresden, MIT-LL, Sandia Labs are some examples}.

\subsection{Industry Partners}
\subsubsection{NHanced}
NHanced Semiconductors has assembled 2.5/3D integrated circuits for more than a decade. 
NHanced has successfully assembled many heterogeneous 3D devices with elements such as memory, FPGAs, microprocessors, ROICs, and sensors. They have shown active circuits stacked to a height of 8 layers. NHanced has produced dozens of different 3D devices and several Focal Plane Array (FPA) and Logic on Memory (LoM) parts for customers. 3D assembly of test wafers has been demonstrated to a 3D stack height of 16 layers.  

To date, NHanced has worked with more than 100 groups to produce 3D integrated circuits.
NHanced performs the 2.5/3D assembly process in its own facility in North Carolina.
NHanced 2.5D and 3D assembly technology covers several techniques including copper to copper diffusion bonding and covalent oxide bonding. NHanced can perform these and other assembly techniques in both wafer-to-wafer and die-to-wafer assembly. Correct operation and reliability of previously 3D assembled devices has been verified at temperatures down to 77\textdegree~K. 
The DBI process has been demonstrated with maximum processing temperature as low as 125\textdegree~C.

\paragraph{NHanced DBI Processing}
A key step in the DBI process is the direct bonding of two surfaces (dies or wafers) using Van der Waals forces. The direct bond process is a two-stage kinetic reaction. At room temperature, the wafers are aligned and brought into contact. 
The wafers have a natural dipole layer of moisture (water) adsorbed onto the surface that initially repels the opposite surface. 
The wafers “float” above each other until a gentle force, usually a mechanical pin, presses lightly at the wafer edge. 
This overcomes the electrostatic repulsion so that the water molecules bond through Van der Waals forces. Subsequent heating to high temperatures is necessary to achieve high bond strength through the formation of covalent Si-O-Si bonds.
This reaction requires a high thermal budget that is not suitable for the fabrication of many devices.
However, modifying the surface chemistry allows the formation of chemical bonds at much lower temperatures. 
Such surface modification technology has been reported and patented. 

One version simply requires a plasma treatment followed by an aqueous ammonium hydroxide rinse. 
A second method first prepares a metal structure(s) interconnecting to a surface (via) or backside (TSV) metal, then deposits oxide, followed by Chemical-mechanical polishing (CMP) to expose the metal/SiO$_2$ co-planar surface. This method is a single mask level process if the seed layer used for electroplating is blanket-etched following electroplating. 
These unit process steps are similar to those used in volume foundry interconnect stack fabrication manufacturing. This method forms the basis of the DBI process.

In the case of copper, special CMP techniques must be used to eliminate the dishing that normally occurs in the soft copper metal in the presence of the hard oxide during CMP. Subsequent heating of the bonded structures in a standard clean room oven forms a monolithic, low resistance metal-metal interface.
Typically, a 300°C anneal is used for formation of a low resistance Cu-Cu interface. Since the activated and terminated oxide layers are bonded together with high strength, the M-M interface is subject to sufficient internal pressure so that when the copper expands at elevated temperatures, a reliable metallic bond results, even when the anneal temperature is lower than a standard anneal for Cu. NHanced has produced commercial devices using a maximum processing temperature of 150°C.

\paragraph{NHanced Vertical Interconnect Size and Pitch}
NHanced’s 3D technology stands out from other 3D integration technologies because of its very fine-grained (down to a few micrometers) small pitch vertical interconnect capability. 
NHanced has assembled devices with more than 10 million vertical interconnects per layer and has demonstrated 3D assemblies with up to 8 device layers.
NHanced has developed and fabricated a number of different functional devices using this process. 
The fabricated devices have passed preliminary qualification, including –65 to +150°C temperature cycling. Most sensor development has been on 180~nm and 130~nm processes, with some work on 65~nm and 55~nm technologies. 

\subsubsection{Cactus Materials}
Cactus Materials is located in Tempe AZ within Macro-Technology Works (MTW) at Arizona State University Research Park. It has office and lab facilities including class 100 cleanroom located in one of the best semiconductor facilities in Arizona originally built by Motorola.

\paragraph{Capabilities}
Cactus Materials, Inc. has in-house laboratory facilities for surface activation, wet processing and chemistry, wafer bonding and alignment equipment, and characterization tools. Solvent wet processing, in-situ surface activation in vacuum environment, wafer bonding and alignment, as well as IR interface void characterization can be performed at Cactus.

Cactus Materials, Inc. is 10 mins away from ASU Nanofab where Cactus has direct access to state-of-the-art semiconductor equipment at discounted hourly rates. PI and engineers from Cactus Materials are trained and qualified to run those tools and processes.

ASU Nanofab has STS AGE for plasma system, PECVD, resist coater/spinner/developer, lithography, thickness measurement profilometer and electrical testing and C-V measurement.
Cactus Materials, Inc. has a long term agreement (five years) with ASU Eyring Materials Center \footnote{http://le-csss.asu.edu} for access to SEM, TEM, Ion Beam Analysis, Auger Electron Spectroscopy (AES) which will be used to characterize bonding interface. An agreement can be provided upon request. The PI has experience in performing TEM characterization, and an engineer at Cactus Materials will run SEM at ASU.

\section{Foreseen applications for 3D integration}
\label{sec:applications}
There are a number of currently active 3D R\&D projects for High Energy Physics and 
related fields. These include single-photon avalanche diode 3D integration on CMOS\cite{7838372}\cite{8617983},
3D integration of Medipix and Timepix\cite{Medi}, TSV insertion into existing chips, X-Ray imaging\cite{Fahim:2017vno}. and several  others\cite{Yamada:2019ule}\cite{Zhang:2020rgo}. The Timepix 4 
collaboration has recently demonstrated a 4-side buttable chip 
utilizing TSVs to achieve $>$99.5\% active area in a bump-bonded array
\cite{Llopart:2022hyz}. In the following 
sections we comment a few potential applications in HEP.

\subsection{High granularity LGADS}
The recent development of silicon diode Low-Gain Avalanche Detectors (LGADs)~\cite{bib:LGAD,bib:UFSD300umTB} has enabled the design of granular ($\thicksim$1x1 $mm^2$) fast-timing layers for the ATLAS and CMS tracking systems at the HL-LHC. 
These systems will allow the determination of the time-of-passage of minimum ionizing particles to a precision of better than 50~ps~\cite{Collaboration:2623663}. 

The essential design aspects of the LGAD can be described as a ``p++''  with a dopant concentration significantly greater than that of the ``p'' bulk. 
This leads, after depletion, to an electric field large enough to provide amplification (by as much as a factor of 70) through multiplication of the signal. 
Because of this amplification, the ``p'' region can be made very thin (50~$\mu m$ or less), leading to a fast signal and, in turn, precise timing. 
Standard LGADs are limited in terms of granularity, however several new technologies are being studied to overcome this limitation.
These technologies, when refined, will allow to produce finely segmented LGADs down to less than 50~$\mu m$ scale maintaining the exceptional time resolution.
The ultimate goal is to provide readout to LGAD sensors with the ability to reach 10~$\mu m$ of position resolution and 10ps of time resolution.

\subsection{Double Sided and Small Pixel LGADs}
The basic concept of the double sided LGAD consists of a double-sided silicon detector with
a gain layer on the electron-collecting side and an array of small (3D-integrated) electrodes
on the hole-collecting (anode) side~\cite{osti_1841398}. We assume that the detector is thick compared to the anode pitch. The electron-collecting cathode will observe a fast rise-time signal due
to the avalanche in the nearby gain layer, while the anodes can provide information on the
depth and location of the charge deposition. This complementary approach allows us to
measure timing with coarse segmentation in the cathodes and therefore lower the total
power and complexity for the timing layer. The anode signal shapes reflect an initial peak
due to the primary ionization and a secondary peak a few nanoseconds later due to holes
generated in the gain layer. We can use the anode signals to either measure total collected 
charge for position resolution or the current pulse shape to measure the depth of the charge
deposition at the pixel position.

\subsection{3DIC SiPM}
The Silicon Photomultiplier (SiPM) has become a staple photodetector for high energy
physics, replacing the photomultiplier in many applications. Although inherently a digital
device, the current generation of SiPMs rely on analog summing of the micropixels and
resistive quenching of the avalanche. A 3DIC version of the SiPM can incorporate much
more sophisticated processing including active quenching for each pixel, digital timing and
windowing, intermicropixel communication and digital pattern recognition and readout.
Such a device can be tailored to the application and can be more selective and much more
powerful that the usual analog SiPM. Prototypes are under development by the Pratte
group at Sherbrooke University~\cite{Parent:2018buj}.

\subsection{3DIC for high performance Pattern Recognition}
High performance pattern recognition capability will become more important in the future. Traditionally, pattern recognition capability has been implemented in either FPGAs
or conventional ASICs. Adding a “third” dimension opens up the possibility for new architectures that could significantly enhance pattern recognition capability. The 3DIC based
architectures allow massive three-dimensional networks for data communication with much
shorter traces and very low parasitic capacitance, with flexible algorithm cells distributed
throughout the network. With this kind of data communication network, pattern recognition become much easier and one could even mimic the detector structure for pattern
recognition, such as track finding or particle flow over multiple detector layers using both
position information and time of arrival information. The basic algorithm cells could be as
simple as Content Address Memory cell for simple matching, or could be as sophisticated
as NN cells to form a high performance NN network. One simple example~\cite{LIU20121973} is using 3DIC as a way to implement associative memory structures for fast track finding applications.

\subsection{Edgeless Tile Arrays}
Silicon based sensors can be processed at wafer scale, such as the 8” sensors for the CMS High Granularity Calorimeter, but electronic integrated circuits are generally limited by the size of the photomask reticule, typically 2 × 3 cm or less. This mismatch forces us to engineer
complex multi-chip modules where the geometry and functionality of the pixel array is
limited by the availability and routing of the external interconnects. The combination of
3D electronics, which provides dense interconnects and vertical through-silicon readout,
with active edge sensors that limit dead regions normally associated with silicon detectors,
allow us to build fully active tiles. Tiles would be assembled into an array only after they
have been fully tested. The only external interconnect are the backside readout bump
bonds that are bonded to the detector motherboard. Tiles can populate complex shapes
with near-optimal tiling and low dead area. All fabrication processes are wafer-scale, which
lowers the processing costs. 

\subsection{Small Pixel Induced Current Detectors}
3D integration allows for small pixels with minimal capacitance associated with the inter-connect and electronic processing in complex, multi-tier readout integrated circuits. 
The resulting signal/noise can exceed that provided by LGADs, which typically have much higher load capacitance. 
The induced current pulse is prompt, with very fast rise time, and, combined with low capacitance, has the potential to provide excellent time resolution~\cite{LIPTON2019162423}. A fast trans-impedance amplifier coupled to the small pixel should provide both excellent time and spatial resolution. In addition to the time resolution provided by the central pixel, the detector can also utilize the transient currents determined by weighting field coupling to nearby pixels. 
Shapes of these output signals depend on the geometry of charge motion within the silicon with respect to the electrode location. 
The signals shapes can be used to provide track angle information and remove off-angle background tracks. 
The design of such a device can be flexible, a thin detector to optimize radiation hardness, or a thicker detector to provide information on track angle or charge deposition pattern.

\subsection{Zero mass tracker}
3D integration of thin sensors with lightweight readout chips would allow to build low (or almost zero) mass trackers.
A zero mass tracker would be extremely useful in many detectors where the preservation of the energy of the particles escaping from the interaction point is very important.
The thin layers of detectors/readout, after thinning is foreseeable to have a 50~$\mu m$ sensor (or less) plus a 10~$\mu m$ of electronics per layer.
Several thin layers would give a very precise estimation of the interaction point without affecting the energy reading of subsequent sub-detector systems.

\subsection{2D Interconnects and Interposers}
The size and complexity of substrates (flex, PC board, ceramic) for 
HEP applications has become a significant limitation on detector design. 
We often use multi-chip modules to extract signals from front-end sensors, 
process them and send the resulting signals to the readout\cite{Kovacs:2017qxr}. 
Replacing conventional hybrid circuits with silicon-based "2.5D"
substrates can improve the bandwidth, density and power efficiency of these systems. 
A silicon-based substrate or interposer (with TSVs), combined with microbump 
technology connecting the substrate to component "chiplets" can improve 
performance by an order of magnitude or more\cite{9290276}. 
This can allow for  
substantially enhanced on-detector processing or overall bandwidth.

\subsection{X-rays}
This technology would make possible to build stacks of 3D interconnected sensors/readout that can be used for pattern recognition tracking~\cite{LIU20121973} and for X-ray detection applications with thin sensors.
For mid-high X-ray detection a fair amount of material is necessary to achieve a reasonable detection efficiency. 
However a thick enough LGAD would lose all the properties of a thin sensor (good timing, fast repetition rate) and a standard stack of thin LGADs would have a fraction of inactive region.
3D integrated layers of LGAD and readout would have both the fast properties of thin sensors (allowing high repetition rate) and the large active thickness necessary for stopping high energy X-rays.

\subsection{Dense active target}
The stacks of 3D interconnected sensors/readout mentioned in the previous section could also be applied to the construction of high granularity active targets for particle decay reconstruction.
The fast charge collection time allows for great pulse pair resolution, paired with high granularity and low inactive area it would allow to fully reconstruct particle decay chains.
Fast high granularity active targets are of great interest, to give an example, for next generation pion and muon decay experiments (e.g. PIONEER~\cite{Mazza:2021adt})


\section{Results and ongoing projects}
\label{sec:results}
\subsection{Latest FNAL results}
Fermilab has had a long history of R\&D into 3D integrated circuits. 
This began with the stringent requirements imposed by the ILC vertex 
detector. They worked with MIT-LL on an early ARPA-sponsored 3D prototype 
run. They then moved to prototypes with Ziptronix and Tezzaron/NHanced 
utilizing foundry-provided TSVs and an early version of the DBI 
process. This 3D multiproject run - after a long development cycle and 
several iterations - yielded functional two-tier chips for ILC, HL-LHC, and 
X-Ray imaging. These chips were in turn 3D(DBI) chip-to-wafer bonded to 
silicon sensors - yielding a three-layer stack with readout layers
35 microns thick. This allowed a direct comparison of identical ROICs and detectors that were assembled with bump and DBI bonds. The DBI bonded 
assemblies had roughly half ($37.7 \pm 2$ vs $70 \pm 10$) the noise of the bump bonded set\cite{Lipton:2015vca} due to the lower interconnect capacitance. 

During a following R\&D run all three companies involved changed ownership
and TSV capability in 130nm was no longer available. This run was never 
completed.  The ASIC priority on the near future is the completion 
of the LHC upgrades and developments for the neutrino program. Fermilab 
is continuing to work with MIT-LL, NHanced and others on 3D technologies that 
could enable processing of fields of pixels using logic, neural networks, or
machine learning layers on a 3D processing stack. 

\subsection{UCSC+Cactus ongoing projects}
The University of California at Santa Cruz (UCSC) is working in collaboration with Cactus Materials, Inc. to develop a 3D integration process for high granularity LGADs.
The goal is the packaging of fine pitched AC-coupled LGAD (AC-LGAD) sensors from the FBK RSD2 production with a dummy TSV wafer to prove successful interconnection.
Once the technology is developed it will allow to readout finely pixelated AC-LGADs for 4D tracking applications.
Furthermore the UCSC-Cactus collaboration is pursuing device engineering using the company's wafer to wafer bonding capabilities.
One example is the use of W2W bonding for production of devices like the DJ-LGAD~\cite{Ayyoub:2021dgk}.

\section{Path for future development}
\label{sec:future}
The collaboration between national laboratories, universities and industry with established expertise in advanced packaging is crucial for the successful introduction of 3D integration in the high-energy physics research community.
This comes naturally since this technology is much more advanced in industry (some future outlooks in~\cite{3DintArticle}) than it is in research.
Such partnership can provide cost-effective implementations of this technology to use in a research setting.
Advanced packaging will enable the pursuit of 4D tracking to a level of precision beyond what is currently achievable.
Future high-energy physics, nuclear physics and photon physics efforts (as well as many other fields) have to consider advanced packaging and industry partnerships as a possibility to reach their set goals.
This will allow to advance the introduction of this technology in the global research community.
A few US based group are pursuing this goal in collaboration with small volume companies listed in section~\ref{sec:companies}. A short description of the ongoing projects and results can be found in section~\ref{sec:results}. 

\bibliographystyle{unsrt}
\small
\bibliography{bibliography.bib}

\begin{thebibliography}{10}

\bibitem{7838372}
T.~Al~Abbas, N.~A.~W. Dutton, O.~Almer, S.~Pellegrini, Y.~Henrion, and R.~K.
  Henderson.
\newblock Backside illuminated spad image sensor with 7.83um pitch in
  3d-stacked cmos technology.
\newblock In {\em 2016 IEEE International Electron Devices Meeting (IEDM)},
  pages 8.1.1--8.1.4, 2016.

\bibitem{8617983}
Edoardo Charbon, Claudio Bruschini, and Myung-Jae Lee.
\newblock 3d-stacked cmos spad image sensors: Technology and applications.
\newblock In {\em 2018 25th IEEE International Conference on Electronics,
  Circuits and Systems (ICECS)}, pages 1--4, 2018.

\bibitem{Medi}
Michael Campbell, J.~Alozy, Rafael Ballabriga, Erik Fröjdh, Erik Heijne,
  X.~Llopart, T.~Poikela, L.~Tlustos, Pierpaolo Valerio, and Will Wong.
\newblock Towards a new generation of pixel detector readout chips.
\newblock {\em Journal of Instrumentation}, 11:C01007--C01007, 01 2016.

\bibitem{Fahim:2017vno}
Farah Fahim et~al.
\newblock {Vertically Integrated Edgeless Photon Imaging Camera}.
\newblock 2017.

\bibitem{Yamada:2019ule}
Miho Yamada, Shun Ono, Yasuo Arai, Ikuo Kurachi, Toru Tsuboyama, Masayuki
  Ikebe, and Makoto Motoyoshi.
\newblock {3D Integrated Pixel Sensor with Silicon-on-Insulator Technology for
  the International Linear Collider Experiment}.
\newblock In {\em {IEEE International 3D Systems Integration Conference}}, 10
  2019.

\bibitem{Zhang:2020rgo}
Jie Zhang et~al.
\newblock {The TSV process in the hybrid pixel detector for the High Energy
  Photon Source}.
\newblock {\em Nucl. Instrum. Meth. A}, 980:164425, 2020.

\bibitem{Llopart:2022hyz}
X.~Llopart et~al.
\newblock {Timepix4, a large area pixel detector readout chip which can be
  tiled on 4 sides providing sub-200 ps timestamp binning}.
\newblock {\em JINST}, 17(01):C01044, 2022.

\bibitem{bib:LGAD}
G.~Pellegrini et~al.
\newblock {Technology developments and first measurements of Low Gain Avalanche
  Detectors (LGAD) for high energy physics applications}.
\newblock {\em Nucl. Instrum. Meth.}, A765:12 -- 16, 2014.

\bibitem{bib:UFSD300umTB}
H.~F.~W. Sadrozinski et~al.
\newblock {Ultra-fast silicon detectors (UFSD)}.
\newblock {\em Nucl. Instrum. Meth.}, A831:18--23, 2016.

\bibitem{Collaboration:2623663}
ATLAS.
\newblock {Technical Proposal: A High-Granularity Timing Detector for the ATLAS
  Phase-II Upgrade}.
\newblock Technical Report CERN-LHCC-2018-023. LHCC-P-012, CERN, Geneva, Jun
  2018.

\bibitem{osti_1841398}
Ronald~J. Lipton.
\newblock A double sided lgad-based detector providing timing, position, and
  track angle information.
\newblock 1 1900.

\bibitem{Parent:2018buj}
Samuel Parent, Maxime C\^ot\'e, Fr\'ed\'eric Vachon, Robert Groulx, St\'ephane
  Martel, Henri Dautet, Serge~A. Charlebois, and Jean-Fran\c{c}ois Pratte.
\newblock {Single Photon Avalanche Diodes and Vertical Integration Process for
  a 3D Digital SiPM using Industrial Semiconductor Technologies}.
\newblock In {\em {2018 IEEE Nuclear Science Symposium and Medical Imaging
  Conference}}, pages 1--4, 2018.

\bibitem{LIU20121973}
Ted Liu, Jim Hoff, Grzegorz Deptuch, and Ray Yarema.
\newblock A new concept of vertically integrated pattern recognition
  associative memory.
\newblock {\em Physics Procedia}, 37:1973 -- 1982, 2012.
\newblock Proceedings of the 2nd International Conference on Technology and
  Instrumentation in Particle Physics (TIPP 2011).

\bibitem{LIPTON2019162423}
Ronald Lipton and Jason Theiman.
\newblock Fast timing with induced current detectors.
\newblock {\em Nuclear Instruments and Methods in Physics Research Section A:
  Accelerators, Spectrometers, Detectors and Associated Equipment}, 945:162423,
  2019.

\bibitem{Kovacs:2017qxr}
M.~Kovacs, G.~Blanchot, T.~Gadek, A.~Honma, and A.~Koliatos.
\newblock {HDI flexible front-end hybrid prototype for the PS module of the CMS
  tracker upgrade}.
\newblock {\em JINST}, 12(02):C02029, 2017.

\bibitem{9290276}
Saptadeep Pal and Puneet Gupta.
\newblock Pathfinding for 2.5d interconnect technologies.
\newblock In {\em 2020 ACM/IEEE International Workshop on System Level
  Interconnect Prediction (SLIP)}, pages 1--8, 2020.

\bibitem{Mazza:2021adt}
Simone~Michele Mazza.
\newblock {An LGAD-Based Full Active Target for the PIONEER Experiment}.
\newblock {\em Instruments}, 5(4):40, 2021.

\bibitem{Lipton:2015vca}
Ronald Lipton et~al.
\newblock {Three Dimensional Integrated Circuits Bonded to Sensors}.
\newblock {\em PoS}, Vertex2014:045, 2015.

\bibitem{Ayyoub:2021dgk}
S.~Ayyoub, C.~Gee, R.~Islam, S.~M. Mazza, B.~Schumm, A.~Seiden, and Y.~Zhao.
\newblock {A new approach to achieving high granularity for silicon diode
  detectors with impact ionization gain}.
\newblock 1 2021.

\bibitem{3DintArticle}
{Next-Gen 3D Chip/Packaging Race Begins}.
\newblock https://semiengineering.com/next-gen-3d-chip-packaging-race-begins/.

\end{thebibliography}

\end{document}